# CMS Workflow Execution using Intelligent Job Scheduling and Data Access Strategies

Khawar Hasham[1,2], Antonio Delgado Peris[3], Ashiq Anjum[1], Dave Evans[4], Dirk Hufnagel[2], Eduardo Huedo[5],
José M. Hernández[3], Richard McClatchey[1], Stephen Gowdy[2], Simon Metson[6]

1. University of the West of England (UWE), Bristol, UK
2. European Center for Nuclear Research (CERN), Geneva, Switzerland
3. Centro de Investigaciones Energéticas, Medioambientales y Tecnológicas (CIEMAT), Madrid, Spain
4. Fermi National Accelerator Laboratory (Fermilab), Illinois, USA
5. Universidad Complutense de Madrid (UCM), Madrid, Spain
6. University of Bristol, Bristol, UK

{khawar.ahmad, antonio.delgado.peris}@cern.ch

*Abstract*—Complex scientific workflows can process large amounts of data using thousands of tasks. The turnaround times of these workflows are often affected by various latencies such as the resource discovery, scheduling and data access latencies for the individual workflow processes or actors. Minimizing these latencies will improve the overall execution time of a workflow and thus lead to a more efficient and robust processing environment. In this paper, we propose a pilot job based infrastructure that has intelligent data reuse and job execution strategies to minimize the scheduling, queuing, execution and data access latencies. The results have shown that significant improvements in the overall turnaround time of a workflow can be achieved with this approach. The proposed approach has been evaluated, first using the CMS Tier0 data processing workflow, and then simulating the workflows to evaluate its effectiveness in a controlled environment.

*Index Terms*—workflows, latency, pilot jobs, data cache, grid

## I. INTRODUCTION

SCIENTIFIC experiments such as the CMS experiment [1] at CERN, Geneva, produce large amounts of data, which are then consumed by a variety of applications and users around the world. Various forms of scientific analyses, data reconstructions and data derivations are performed on the scientific data. These analyses use workflows to process thousands of files, to execute tasks and to take care of the dependencies between these tasks. Examples of such an analysis include the CMS Tier0 workflows [2] that process the CMS data at CERN. The turnaround time of these workflows depends upon the number of files being processed and the number and nature of the tasks within the workflow.

In a stand-alone environment, the turnaround time of a workflow, running on a single machine, is simply the sum of the execution times of individual actors in that workflow. Using this environment, it would take an enormous amount of time to execute a complete workflow on a single machine because all workflow actors would have to run sequentially. The situation becomes particularly complex and very time consuming if the workflows also operate on large data sets. The problem is further compounded if a number of users submit multiple tasks, each in turn consuming multiple datasets, in order to achieve desired results. However, independent available workflow actors, whose requirements have been met and have no dependencies, can run in parallel in a distributed environment. Therefore, tasks in scientific workflows are preferably executed on distributed resources to reduce the overall execution time and to enable users to achieve rapid throughput.

In the case of a highly distributed environment such as the Worldwide LHC Computing Grid (WLCG) [3], which has been deployed for the analysis of data from the Large Hadron Collider (LHC), each workflow actor would face scheduling and data access latencies during its lifecycle (see Figure 1). The WLCG is a global collaboration of more than 170 computing centres in 34 countries that combines the computing resources of more than 100,000 processors. The mission of the WLCG project is to build and maintain data storage and an analysis infrastructure for the entire high energy physics community that will use the data from the LHC at CERN. At full operation intensity, the LHC will produce roughly 15 Petabytes (15 million Gigabytes) of data annually that thousands of scientists around the world will access and analyse.

Grid scheduling latency is the cumulative time spent in discovering resources in a Grid for scheduling and the waiting time that is spent in the queues of meta and local schedulers before a job can start execution on a so-called worker node (WN). A worker node is an execution resource at a site. Here the data access latencies are mainly caused by the network bandwidth limitations, the load on a Storage Element (SE) and the time spent in accessing a storage media such as a tape drive [4]. These latencies can affect the turnaround time of the workflow and in some cases can exceed the overall execution time of a job. An experimental study [5] has shown that it takes almost five minutes (on average) for a job, in the EGEE Grid [6], to start its execution from the time it was submitted. One can understand the extent of delays if there are thousands of jobs being submitted and executed in a Grid infrastructure such as the WLCG. Minimizing these latencies is a major research challenge in order to offer a high quality of service that users expect from production Grids.

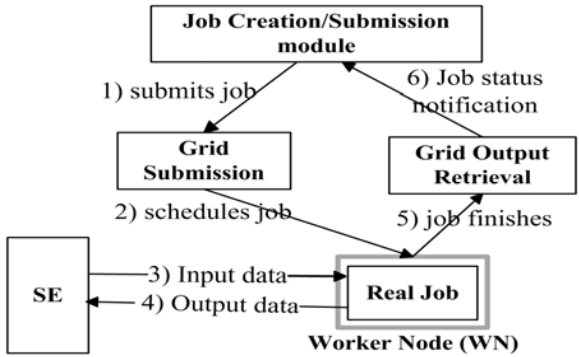

Fig. 1. A Job life cycle in a Grid environment

With the current data storage hierarchy of the WLCG, each site maintains one or multiple dedicated machines called storage elements (SE) to store data. Each job can access the data from a given SE. The jobs in the CMS Tier0 workflow (detailed in Section III) stream data directly from chunks of the data available on the SE. These jobs process this data, without downloading the entire dataset on the local hard disk of a worker node. This mechanism (see Figure 2) creates an additional burden on the SE if every CPU-bound job remotely accesses small chunks of the data periodically leading to a high frequency of I/O requests. An SE has to keep the files open, as they are being read, for longer periods of time and this can add to the latency being faced by the other data requests.

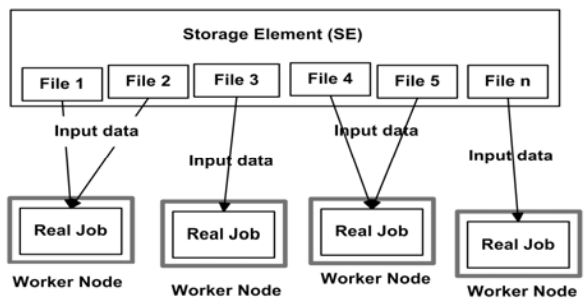

Fig. 2. Multiple jobs accessing an SE

Storage systems such as CASTOR [7] can store petabytes of data, however, such systems are vulnerable to performance issues in terms of high access latencies and this becomes worse with increasing loads. This leads to longer data access times and thus affects the overall execution time of a workflow. In order to reduce these data access and scheduling latencies and to improve the workflow turnaround time, this paper proposes to use a pull-based scheduling system and to establish data caches on the worker nodes. This can be achieved by managing the resources of a worker node by using a customized resource management software component.

To demonstrate this work, the proposed approach makes use of a global scheduler and the concept of a *pilot job*. A pilot job is a job that is responsible for setting up the required execution environment and for managing the execution of a real job. A real job is a job that is part of a user workflow and that waits in the global scheduler queue. Both these jobs follow different submission and scheduling mechanisms. A pilot job follows the traditional grid submission mechanism, however, a real job will bypass it because a pilot job downloads it from a global scheduler queue for execution. With the help of this approach, a pilot job can assist the real job in finding all or some of its required files in the cache maintained on the worker nodes. A real job can start its execution as soon as it has been scheduled to a pilot job thus reducing the queuing and scheduling delays. The real job will first look for its input files in the cache and will read the data from the local cache, provided the cache holds the required data, otherwise the real job will contact the given SE for the data. Once a real job has completed its execution, a pilot job immediately notifies its completion status to the scheduling and monitoring components, thus, minimizing the delays that otherwise exist in retrieving and notifying the job completion status. This approach is further explained in Section IV.

This paper is organized as follows. Section II discusses the state of the art in the research domain. Section III briefly outlines the Tier0 workflow specification and execution system being used at CMS-CERN. This workflow is being taken as a case study to demonstrate that the proposed approach is effective in improving the data access, queuing, scheduling and execution latencies in real scientific computing environments. Section IV provides details of the proposed architecture and justifies its selection in solving the problem. Section V provides a description of the results which show that the proposed solution has been quite effective in reducing the turnaround times of large workflows. Section VI concludes this paper with possible directions for future work.

## II. RELATED WORK

Numerous efforts have been made to reduce data access latencies in intensive data processing applications. The replica management service [8] of the Globus toolkit uses data replication in order to optimize the performance of a data analysis process. The data replication is done at the site level, however, it cannot solve the latency issues resulting from a large number of open file pointers on the SE and a large number of I/O requests. Intelligent Cache Management (ICM) [9] uses the concept of a local data cache to optimize query performance but it replicates and stores the data on a regional basis. None of these approaches exploits the resources at worker nodes for the purpose of data caching. Peer-to-peer (P2P) approaches [10] have been using end node capabilities for data storage, most notably, BitTorrent [11] and super-peer approaches such as KaZaa [12] use end node capabilities for data discovery and data transfer. The BitTorrent approach works on the so-called fair share basis. Data providers have to supply data for consumption by consumers in the outside world, which puts additional burden on the network usage and could also be against the security policies of the Grid sites. Taylor in [13] proposes a framework that uses the concept of super peers to create an application-specific or workflow-specific data cache overlay. This approach makes use of AlChemist's built-in flexibility to support a P2P infrastructure on top of the WSPeer API for communication with its peers. However, this approach is dependent upon the AlChemist framework and the WSPeer API to create data cache overlays on dedicated data nodes, whereas we have proposed to create data cache on every worker node inside a cluster to optimally use the available resources in the Grid infrastructures. There have been various approaches [14] to schedule job taking into account the location of the data needed by a job to minimize the data transfer, and thus the job execution time.

In addition to these efforts, research has been carried out to minimize job submission and output retrieval latencies by using the concept of pilot jobs in Grids. Grid projects such as PanDA [15], DiRAC [16] and AliEn (Alice Environment) [17] use this approach to schedule and execute real jobs. All these projects use the pilot jobs to reduce the job submission latency by pulling a job from a global job queue and thus provide a fault-tolerant job execution mechanism. However, these systems do not use a pilot job infrastructure to reduce the data access latencies. A project in CMS, GlideInWMS [18], makes use of grid resources as part of its Condor pool. It uses Condor [19] glidein which acts as a pilot job on a worker node. It takes the leverage of the Condor infrastructure to enable communication with different Condor daemons. Since these glideins are often running behind a firewall, it uses a workaround called Condor's Generic Connection Brokering (GCB) [20] which helps the global scheduling daemons to contact these glideins and to push the actual jobs directly to them. However, this approach has led to scalability problems [21]. Moreover, it does not support the data cache mechanism on worker nodes to reduce data access latencies.

The work done by Shankar et al. [22] is closely related to the work being reported in this paper. Their approach makes use of a dedicated cache space on the worker nodes in an execution cluster for the data caching purpose. They accomplish this with the help of condor-DAGMan, which makes it specific to the Condor environment only. Its scheduling process involves prior planning of the resources for a given DAG, however, in environments such as CMS, jobs are generally data driven and are not completely known until they have been created. Moreover, the scheduling is performed within a single site and hence is not suitable for heterogeneous environments like the WLCG Grid.

III. CASE STUDY

The CMS experiment at CERN uses a multi-tier distributed architecture [23] where CERN is the Tier0. Using a four-tiered architecture (from CERN's central computer as the Tier0 to small Tier3 analysis clusters), CERN distributes LHC data and computations across resources worldwide to achieve aggregate computational power unprecedented in high energy physics data analysis research. The Tier0 reformats, writes out Primary Datasets, and stores this raw data, generated from the output of the CMS Detector, performs an initial data reconstruction and distributes the processed data to Tier1s. In this paper we concentrate on the Tier0 workflows and their data access patterns, however, the approach being discussed in this paper should be of wider usability, especially for other CMS data intensive workflows that we intend to demonstrate in future. For the initial data reconstruction, a Tier0 workflow is used, which is also a sample workflow to evaluate and benchmark the proposed system. This workflow has three main steps, namely 1) *Repacker* 2) *PromptReco* and 3) *AlcaReco*. The *Repacker* jobs reformat the binary data from CMS detector and split the output into different Primary Datasets based on physics information. The *PromptReco* jobs take this output as their input and perform an initial reconstruction into usable sets of physics data such as the particle trajectories and the properties of the candidate particles. The *AlcaReco* jobs perform much higher selectivity of the data produced by the *PromptReco* jobs and also carry out some processing on this small subset. This output is used to align and calibrate the CMS detector. Figure 3 shows the CMS Tier0 workflow.

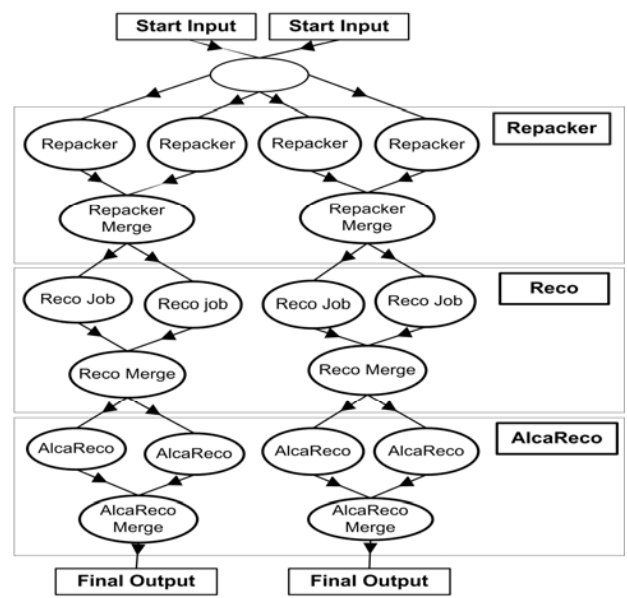

Fig. 3. Tier0 workflow for CMS at CERN

In each step, several jobs are created. The number of jobs in each step depends on the number of physics events (or filtered particle collisions of interest) in the input files. Currently each job has to process around 5000 CMS physics events. Each job produces a relatively small output data as compared to its input data. It is inefficient to store and transfer smaller files to a tape-based central storage system because the process encounters delays and latencies in transferring a file to and from the tape drives. Therefore, each step has a special job called the *Merge* job, which gets the output from multiple jobs and merges them. Only the merged files should, ideally, be transferred to the central storage system in the first instance.

The creation and execution of all the workflow actors is data driven. The workflow starts execution whenever a new file is available that requires some kind of processing. The unnamed oval process in Figure 3 triggers the first step by creating the *Repacker* jobs. The subsequent jobs are created according to the system policies, workflow rules and data availability. There are two main characteristics of this workflow. Multiple jobs are dependent on a single input file, and a single job, which is the *Merge* job, is dependent on multiple smaller files produced by earlier jobs in the workflow. This workflow is created and executed by a ProdAgent [24]; a workflow management system used in CMS. ProdAgent is a component based system driven by an asynchronous and persistent messaging system for communication among these components. ProdAgent is responsible for creating, submitting and then monitoring the real jobs in a CMS workflow. In the existing setup, all jobs within a Tier0 workflow are queued up in the global scheduler of ProdAgent. The global scheduler can schedule the jobs on the available sites in the Grid using the configured submission mechanism such as gLite [25] and Condor-G [26]. The Tier0 instance uses local LSF [27] submission.

Once a job has been scheduled from the meta-scheduler, it comes to a local scheduler such as LSF, PBS [28] or Condor running on a particular site. A job has to wait in the local scheduler's queue before it is scheduled to a worker node. After arriving on the worker node, the data dependent jobs undergo a further wait before their required datasets become online on the given SE for streaming. Once the job can access the data, it reads data in chunks and performs its processing. After completing the processing of this data, the job stages back the output to a given SE. It then faces further delays until a monitoring component knows that a job has been finished and it has staged back its output. The latency in retrieving the job completion information delays the submission of a dependent job, thus increasing the workflow turnaround time. In the current execution environment, as shown earlier in Figure 1, each job has to face the afore-mentioned scheduling, monitoring and data access latencies. These latencies affect the execution time of an individual job, which, in turn, affects the turnaround time of the whole workflow.

The CMS Tier0 is a latency critical system, where disk buffers fill up if the data coming from the detector are not processed in a timely manner, and calibration constants derived promptly are used to reconstruct the new data. Therefore, removing aforementioned latencies is very important to improve the turnaround time of the workflow.

## IV. PROPOSED ARCHITECTURE

In order to optimize the execution of the CMS data processing workflow, we propose to use a pull-paradigm driven by pilot jobs and to establish data caches on the execution resources. This approach will help in avoiding scheduling, monitoring and data access latencies for the real jobs. As a result of this approach, there will be fewer job failures that may appear due to incorrect job execution environments. The approach will also be used to create data caches on the worker nodes.

The pilot job concept provides three main advantages. Firstly, real jobs do not face scheduling and monitoring latencies since the pilot jobs will pull them directly from the global scheduling queue and notify job completion as soon as a job has been finished. Secondly, the pilot job will manage the available resources on the worker node for data caching, which will help in avoiding data access latencies. Thirdly, the pilot jobs will ensure that an execution environment is appropriate for a real job before executing it. Furthermore, the pilot jobs act as a layer on top of the local batch system such as Condor and LSF and therefore it ignores the local schedulers and makes use of the meta-scheduler policies for making scheduling decisions. This not only saves the queuing times that can be quite high in local schedulers, but it will also reduce the job failures. As a result of this, jobs are only sent to a site if they are requested by a pilot job running on the site and it has the required execution environment. Moreover, this approach makes the decision making process distributed, cooperative and fault tolerant. With this approach, there will potentially be a single scheduler in the Grid for the real jobs since they will bypass the local schedulers running on the sites. The meta-scheduler in association with the pilot jobs will make cooperative scheduling decisions to reduce job failures and minimize queuing and execution latencies. This proposed approach dynamically matches real jobs to the pilot jobs and thus makes the scheduling decisions that are required for efficient cache and resource usage. The overview of the proposed architecture is shown in Figure 4.

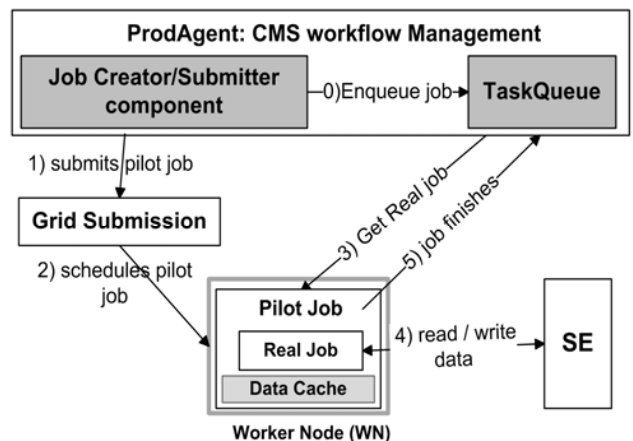

Fig. 4. Overview of the proposed architecture

The *JobCreator* component of the ProdAgent system will create the real jobs from the workflow and enqueue them in the *TaskQueue* (TQ). The *TaskQueue*, a central job queue, will hold all the real jobs of the workflow that are waiting to be scheduled for execution. The *TaskQueue* will schedule them upon receiving job requests from the pilot jobs. The *TaskQueue* is also responsible for registering new pilot jobs and maintaining the information about them. An architecture of the pilot job and the *TaskQueue* is given in Figure 5.

The number of pilot jobs that should be submitted to a site is subject to the number of real jobs that are waiting in the *TaskQueue* for that particular site. Currently, each pilot job is capable of running a single real job at any point in time. Since the Grid sites are shared among multiple Virtual Organizations (VOs), we cannot load them with pilot jobs that will not have work to do. Two configurable thresholds are used to avoid this problem. These thresholds are called *minPilots* and *maxPilots*, which put a limit on the minimum and maximum number of pilot jobs for a site. Each site has its own values for these thresholds that are provided by the site policy. The *PilotMonitor* component, which is responsible for monitoring the state of submitted pilots, calculates the required number of the additional pilot jobs within these thresholds and then requests the *PilotManager* component to submit them. Section IV-A details the algorithm used in the PilotMonitor to calculate the required number of pilot jobs.

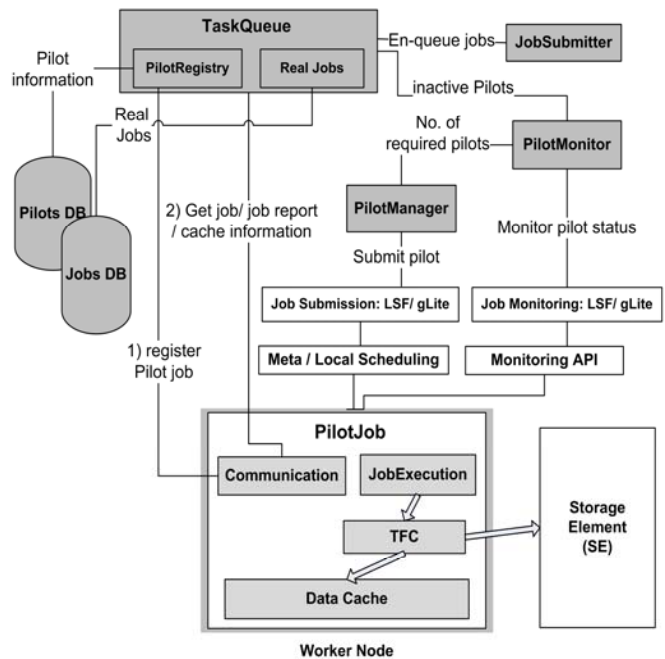

Fig. 5. Detailed architecture of PilotJob and TaskQueue

Upon receiving the request from the *PilotMonitor*, the *PilotManager* component prepares the required number of pilot jobs with configurable parameters and submits them. The pilot jobs are submitted using the underlying submission system such as LSF, Condor or gLite for grid submission. Once a pilot job has been scheduled on a worker node within an execution cluster, it will perform initial environment checks and register itself with the *TaskQueue*. In the registration phase, the *TaskQueue* assigns a unique id, *PilotID*, to each pilot job to identify it during its subsequent requests. Once the environment has been setup and the registration process has been completed, the pilot job is then ready to contact the *TaskQueue* to get the real job. However, if there is something missing in the environment that is required for executing a job, the pilot job announces the error and terminates itself; hence no real job would be executed. This helps in having fewer real job failures that occur due to an improper execution environment which is one of the major reasons for job failures in Grids [29]. The pilot job approach being proposed in this paper will help in reducing such failures.

After the successful environment check, the pilot job contacts a given *TaskQueue* URL and requests for a real job. Section V gives a brief account of the cache-aware scheduling that the *TaskQueue* performs upon receiving the pilot job request. Once a job has started its execution, it looks for the physical location of its input files. The pilot job maintains a mapping file called Trivial File Catalog (TFC) to discover the input files. This is an XML file which maintains the rules to convert a Logical File Name (LFN) into a Physical File Name (PFN) to locate a file. The TFC first looks into the pilot job's cache area for the required files. It provides a pointer to the input file residing on an SE if the required file is not available in the cache. This working is shown in Figure 5.

The pilot job mechanism using the job pull-paradigm is quite useful because it does not pose security concerns for the grid resources. Sites are normally reluctant to open ports to allow the outside world to make connections with their internal resources. The pilot jobs act as clients for the *TaskQueue* and hence address the site security requirements.

*A. PilotMonitor algorithm*

The *PilotMonitor* component keeps track of the submitted pilot jobs and the real jobs enqueued in the *TaskQueue*. The pilot jobs that are submitted to a site can be in one of three states (*inactive, idle, busy*) during their lifecycles. The *inactive* state is applied to those pilot jobs which are not running and have been waiting in the site scheduler. A pilot job will be monitored as *idle* if it is running on a WN but could not get a real job from the *TaskQueue*. A *busy* pilot job means that it has acquired a real job and this is in execution.

The *PilotMonitor* algorithm uses three important thresholds to calculate the required number of pilot jobs for a site. These thresholds are the maximum and minimum number of pilots to be submitted to a site and the minimum number of idle pilots. These thresholds are represented as *minPilots*, *maxPilots*, and *minIdlePilots* respectively. This algorithm makes sure that the required number of the pilot jobs should not exceed the *maxPilots* threshold and also they should not be less than *minPilots*. The last threshold, *minIdlePilots*, may be useful for sites like Tier0, where it may be desirable to always keep some idle pilots that are ready to accept a real job. This minimizes the delay caused by the pilot job submission and also reduces the submission time for the real jobs that are submitted for the first time. All these thresholds are configurable, according to the site policy.

The *PilotMonitor* runs this algorithm periodically for every known site in its list. The algorithm is summarized as follows.

---
**PilotMonitor algorithm**
---

1. Recall thresholds and previously submitted pilots for site
2. Set: available slots = *maxPilots* threshold - submitted pilots
3. If (available slots <= 0)
4.     Then: Do not continue (do not submit more pilots)
5. Query TaskQueue about tasks that can run on this site
6. For each group of enqueued tasks:
7.     If (enqueued tasks < inactive pilots)
8.         Then: mark inactive pilots as active, mark tasks as covered
9.         Else: If (available slots > number of tasks)
10.             Then: send more pilots, mark tasks as covered
11. if (idle pilots < *minIdlePilots*):
12.     Then: send more pilots
13. if (submitted pilots < *minPilots*):
14.     Then: send more pilots

---

This calculation is then passed to the *PilotManager* component which submits the pilot jobs to a given site.

*B. Cache replacement algorithms*

On a worker node, each pilot job will have limited space available for caching so an efficient caching replacement algorithm is required for managing the cache on worker nodes. There are many caching algorithms [30] that can perform this task including the traditional algorithms such as First-In-First-Out (FIFO), Least Recently Used (LRU), and Least Frequently Used (LFU). The traditional algorithms offer low overhead as they need minimal information, such as reference count and last access time, for their cache replacement policies. Here reference count means the number of times a file has been accessed in the past and the last access time means the time at which a file was last accessed. Some improvements have been made in these classical algorithms namely LFU-*, LFU-again, LRU-K [31], and LCB-K [32]. These improved algorithms such as LFU-* remove the cache pollution problem faced by LFU. The LCB-K and other cost sensitive cache algorithms [30] consider the cost of data removal from the cache. However, these improved algorithms store extra information to deal with issues that occur with the traditional algorithms.

The nature of the CMS Tier0 workflow favours the LRU algorithm because once a step has been completed and its output has been merged, the smaller files are no longer required in the following steps. These smaller files are only required by the jobs that were generated at the same level in the workflow hierarchy. The jobs in the following steps use the data from the merged output that has been produced from the smaller files in a previous step. Moreover, the jobs in the CMS workflow do not directly interact with the pilot job's caching component for a cache lookup because they use the TFC to locate the physical location of a file. For these reasons, it is somewhat difficult for the cache component to maintain a reference time history or the reference count, used by the LFU, for each file in its cache. Consequently, for our prototype implementation, we have used LRU because of its simplicity and its compatibility with the CMS workflows.

*C. Data Caching policy*

A pilot job running on a worker node can control resources for the time it is allowed to run. Each real job, running within the pilot job, will consume some input files and generate some output files. Apart from executing a real job, the other important task of a pilot job is to maintain these data files in its cache. The caching policy must adhere to the requirements and constraints detailed in the following paragraphs.

Each running pilot job will be given a certain amount of disk space. The pilot job uses this space to download real jobs and maintain output files. This space will become the pilot job's cache area. This space is configurable at the site level and this is decided by the site administrator. In CMS, each job is given a 10 GB space on the disk. Each pilot job will also get at least 10 GB space that acts as the maximum allowed space for the data storage. Since the jobs are executed within the pilot job space, as shown in Figure 5, we will always need a minimum space available at any given time. This minimum space is used by the real job to temporarily store its output that has been produced from the job execution. Let us call this required minimum space a $Min_{Threshold}$. The total space that can be utilized for caching data can be given as:

$$Cache_{Size} = Max_{Space} - Min_{Threshold}$$

This ensures that we always use the maximum allowed space for caching purpose by always keeping the minimum available space for the job execution.

Let us say we have a set F of 'n' cache files {f1, f2, f3…fn} each having the sizes {S1, S2, S3… Sn} respectively such that their collective sum is less than or equal to cache size

$$\sum_{i=1}^{n} Si \le Cache_{Size}$$

For example, a job produces a new file X which is required to be placed in the data cache. The file X would become part of the cache if the required space is available. If the remaining space in the cache is insufficient to accommodate this new file, then we need to remove some files from the cache. The LRU algorithm should remove files from the cache such that the sum of the removed files matches the following criterion.

$$\sum_{i=1}^{m} Si \ge RequiredSize$$

Where *RequiredSize* is the size of the new file for which the cache replacement algorithm will create space in the cache.

*D. Cache sharing among pilots on same Worker Node*

On execution if resources are available at Tier0 at CERN, multiple jobs can run in parallel on a single WorkerNode (WN). Therefore, it is possible that multiple pilot jobs may land on the same WN. The usage of the cached data will become more effective if these pilot jobs can share their cached data. Since the cached data is available and accessible locally, there will be low data access latencies if the jobs can access the shared cache. The cache sharing concept becomes even more helpful in the scenario when Pilot1 is running a job which needs a file available in the cache of Pilot2 that is running on the same WN. A real job does not need to access an SE if pilots can locate and then share this cached data. This will also increase the cache hit rate.

We propose an approach that is called *cache-per-host* to establish cache sharing among the pilot jobs running on the same WN. Here we assume that the pilot jobs share the same file system on a WN. In the case of CMS, all the pilot jobs run under the same user id or the users belong to the same group, therefore they can access each other's directories. When a pilot job arrives on a WN and registers itself with the *TaskQueue*, the *TaskQueue* sends back the list of other available pilot jobs on that WN and their cache locations in response to this register request. The pilot job will then save this list and poll the given locations for new cache files. A Unix hard link to a newly found file is created into the pilot's own cache area and the file is placed into the cache by using the LRU algorithm that has been discussed previously. In this way, the file remains in the system even if the original owner of that file deletes it. A file is removed from the system only if its last link is deleted. At this point, a pilot job that prompted the file delete operation will notify the *TaskQueue* about this. In *cache-per-host*, the total space available to a pilot job on a worker node for data caching is dynamic. It is calculated as a function of the number of pilot jobs on that worker node, the maximum space allowed to each pilot job and the $Min_{Threshold}$. The following equation shows this model where $num\_pilotjobs$ is the number of pilot jobs on that worker node.

$$Cache_{Size} = Max_{Space} \times num\_pilotjobs - Min_{Threshold}$$

Since the pilots can shut themselves down or new pilots can arrive on the same WN at random, a mechanism is required to update the running pilot jobs about the other available pilot jobs on a particular WN. This is achieved by making use of the 'Heartbeat' message, which a pilot job regularly exchanges with the *TaskQueue*. This message informs the *TaskQueue* that a pilot job is alive. In response, the *TaskQueue* provides the pilot with an updated list of other pilot jobs on the same WN. In this way, each pilot job updates itself about every other pilot job running on the same WN. When a pilot job polls the given pilot jobs' locations if that location is not accessible, then the pilot job removes that entry from its list and that particular pilot job is assumed to be dead. Each pilot job will update its list of the pilot jobs after each 'Heartbeat' message.

*E. Cache sharing among pilots on same Worker Node*

Each job placed in the *TaskQueue* provides its requirements, such as its preferred site and input files. When a pilot job submitted to a worker node starts execution, it will contact the *TaskQueue* to get a job that meets its requirements. The request to the *TaskQueue* includes its *PilotID*, Host, SE, Time-to-Live (TTL) and cached files. In this request, *PilotID* is the id assigned to each pilot job during its registration with the *TaskQueue*, Host is the name of the worker node where the pilot job is running, SE is the name of the storage element accessible to the pilot job in that particular site and the cached files are the files available in the pilot job's cache. The TTL is the remaining life of a pilot job. In the current implementation for the Tier0, the pilot jobs can run forever because resources are dedicated to Tier0 operations. But this information will be configurable in future implementations and will be added into the job scheduling process.

The *TaskQueue* performs the job scheduling by comparing job requirements against the pilot job information. The scheduling algorithm must schedule a job to a pilot job whose maximum requirements meet the information provided by the pilot job. The caching information is used to match job data dependencies against the files maintained by the pilot job. The *TaskQueue* schedules a job to a pilot job that has the maximum number of jobs required files in its cache. A job, arriving on a pilot job that holds some of the required files in its cache, will face less data latency since it can find some or all of its required files in the pilot job's cache. The job without any specific requirement can be scheduled onto any pilot job.

In order to provide improved job scheduling and to use cache more effectively, we implemented a *waitForData* policy alongside the above mentioned scheduling model. According to this policy, when a pilot requests a job but cannot match the data dependencies of a job, the *TaskQueue* would not schedule the job to the pilot if there are other idle pilots holding the required data. The *TaskQueue* would wait for these idle pilots to eventually request a job. In this way, the scheduling process encourages the maximum number of reads from the cache. However, if there are no other pilots that hold the required data or they are not idle, the *TaskQueue* will schedule the job to a pilot that does not have the required files instead of keeping the job for an unknown period of time, because, as a last resort, a job can always access data from an SE.

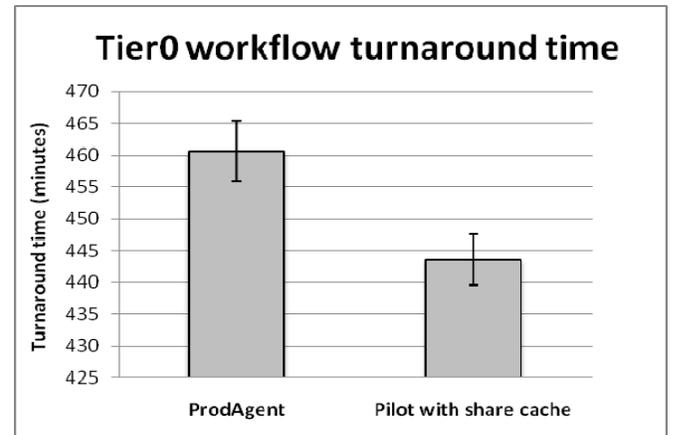

Fig. 6. Effect of pilot-based system on the Tier0 workflow turnaround time

V. EXPERIMENTAL RESULTS

A series of experiments have been conducted at CERN's

Tier0 infrastructure. For these experiments, a test bed has been used that comprises a cluster of 10 machines, each of which is capable of running four jobs in parallel. We used a dedicated resource in Tier0 as an SE to avoid any external influence on the SE. The CMS Tier0 reconstruction workflow is used as a sample workflow in these experiments. This workflow generates a total of 172 jobs, requires 83.41 Giga bytes (GB) of input data, and produces 112 GB of output data. Several iterations of the complete Tier0 workflow have been executed with the existing system i.e. ProdAgent and with the new developed prototype based on pilot jobs and cache. These experiments have been repeated several times. The figures show the measured average values and the error bars represent the standard deviation. The results in Figure 6 show that the workflow turnaround time has been significantly reduced by using the proposed system.

This reduction in the turnaround time is mainly due to the reduction in job submission and job status notification times since the pilot-based approach reduces the job scheduling latencies (explained in the discussion of Figure 7). In these tests, it was not possible to measure the behaviour of the proposed system against different parameters such as job failure rates, queuing times and data access latencies. This is mainly due to the fact that there was no additional load on the SE as it is only being used for data access operations in these experiments. It was not practically feasible to artificially alter the access conditions on the SE that has been used in these experiments. Therefore, a variety of simulation experiments were conducted at CIEMAT (in Madrid, Spain), which is a CMS Tier2 site, to evaluate the impact of the pilot jobs and their data caching patterns.

For the simulated experiments, a simulation engine has been implemented to emulate the ProdAgent and the data driven behaviour of CMS workflows by using a concept called '*steps*'. A workflow is divided in such a way that jobs in the next step depend on the output produced in a previous step. Three types of workflows, generating the jobs in two steps i.e. *step0* and *step1*, have been used in these experiments. These three types of workflows display three different characteristics of data intensive scientific workflows in general and the CMS Tier0 workflow in particular. As mentioned in Section IV, the jobs can display various types of data dependencies. It can be a one-to-one (serial chain) dependency, or many-to-one dependency where one job (the merge job) consumes the files produced by two or more jobs in the previous step, or one-to-many dependency where multiple jobs (splitting jobs) can consume the files produced by a single job.

The serial chain workflow (abbreviated as W1) demonstrates a one-to-one dependency. In this workflow (W1), 80 jobs that produce 80 files as their outputs are created in step0. This is followed by another 80 jobs in step1 that are dependent on the output produced in step0. This workflow represents a one-to-one dependency between the jobs in the workflow. In the second workflow (W2), 40 jobs are created in step0 that produce 40 files and are followed by 80 jobs in step1. In the second workflow, two jobs in step1 are dependent on a single file produced by a job in step0. This workflow represents a splitting workflow where more jobs consume the data that has been produced by fewer jobs in the previous steps. In the third workflow (W3), 80 jobs in step0 produce 80 files and are followed by 40 jobs in step1. This is an example of a merging workflow where two or more than two jobs are merged in the subsequent steps of a workflow. Each job in these workflows produces a file of size 700 Mega bytes (MB). In each workflow, the jobs in step0 are first generated and scheduled, and then the jobs in step1, which depend on the data produced by the jobs in step0, are generated and enqueued in the *TaskQueue*.

In order to study the effect of the proposed approach on different type of workflows under different SE conditions (given in Table 1), two different parameters, the *delay factor* and the *failure rate,* are used for these experiments. The *delay factor* is a delay that a job bears in accessing an SE. It is used to simulate the delays, which occur due to the load on an SE, in reading and writing processes. A higher delay factor means longer times are being taken in reading and writing to the SE. The values for *delay factor* used for these simulations are 0.01, 0.15, and 0.50 and are represented as d1, d2, and d3 respectively. Since the CMS jobs keep on reading the data during their entire execution time, we used a Gaussian distribution to measure and represent the data access times at different stages in the job execution process. The other factor, *failure rate*, is used to simulate the probability of failure in reading or writing data to an SE which eventually means failure of a job, hence, it may have a negative effect on the workflow execution. The values for failure rate used are 0, 0.03, and 0.1 and are represented as f1, f2 and f3 respectively. A higher failure rate means higher chances of failure in reading and writing data from and to a data source. Different combinations of these two factors give us different load conditions on an SE. The d1f1, d2f2, and d3f3 combinations represent *Low*, *Moderate*, and *High* loads on an SE respectively. The *Low* load on an SE means that there are not too many read and write requests to the SE; therefore, jobs would not face long data access delays. The *Moderate* load on an SE means that there are a reasonable number of read and write requests to the SE and jobs might face slight delays in reading or writing files. The *High* load means that there are a huge number of requests pending for reading and writing the data to the SE, consequently, the jobs will face longer delays and a higher probability of failure. Table I summarizes these combinations.

Table I. Combination of delays and failure factors

| Combination | Delay factor | Failure factor | Load on SE / SE condition |
|---|---|---|---|
| d1f1 | d1=0.01 | f1=0 | Low/Normal |
| d2f2 | d2=0.15 | f2=0.03 | Moderate/Medium |
| d3f3 | d3=0.50 | f3=0.1 | High/Worse |

A third factor that can influence the experiments is the caching scheme used in an experiment. The effect of the data caching on such environments (such as in CMS) is more prominent since this can significantly influence the overall execution time. Different cache schemes such as *cache-per-host* (C1), *single-pilot-cache* (C2) and *cache-per-host* without *waitForData* logic (C3) have been used in these experiments. In the single-pilot-cache, the pilot jobs running on a WN do not share their cache data with each other. In the cache-per-host, the pilot jobs on a WN can discover and share cache data with each other. For C1 and C2, *waitForData* logic (as discussed in Section IV-E) is active in task scheduling process. In the following figures except Figure 12, the C1 cache scheme has been used.

In order to study the effect in workflow latency in job submission and scheduling, three different job submission mechanisms have been used in these experiments which are 1) direct submission (noTQ), 2) job submission with already running pilot jobs and 3) job submission by submitting the pilot jobs on demand using the *PilotMonitor*. In the first submission mechanism, the *TaskQueue* and the pilot jobs are not used. The jobs are submitted directly to the Grid using the gLite software. In the second submission mechanism, 120 predefined pilot jobs are already running before the new jobs are enqueued into the *TaskQueue*. In this case, the pilot jobs are ready to acquire new jobs and execute them. In the third submission mechanism, the pilot jobs are submitted on demand using the *PilotMonitor* algorithm explained in section IV-A.

The following paragraphs detail the results of the experiments that have been performed using the experimental setup discussed in the previous paragraphs. In order to measure these results, simulated experiments have been repeated several times and the figures present the measured average values and standard deviation is shown as error bars. The plot in Figure 7 shows the number of running jobs over time for a W3 workflow where the jobs were submitted using the three submission mechanisms. There is an initial job submission delay for the direct (without the pilot jobs and the *TaskQueue*) and *PilotMonitor*-based job submission. This delay is due to the scheduling latencies introduced by gLite, and pilot jobs have to wait in a local scheduler's queue before they can run and request the real jobs. However, this is not the case when the pilot jobs are already running, and thus there are no submission delays as the pilots are already waiting for the real jobs. There are no queuing delays for the real jobs since the pilots pull them as far as they can to meet the jobs' requirements.

The results show a decrease in queuing times for the jobs and scheduling latencies when a pilot-based system is used in comparison to the direct submission. In the direct submission mechanism, there is also a huge delay between the time step0 jobs complete their execution and the jobs in step1 are submitted (group of running jobs as shown on the right side of the plot). This is due to the latency introduced by gLite in notifying the job completion. On the contrary, there are almost no such delays between these steps with the pilot-based approach. The knees in Figure 7, for the pilot-based approach, are due to the delays in the submission of step1 jobs after the jobs in step0 have been completed. Figure 7 shows how job submission, scheduling and job notification delays can be reduced using the pilot-based approach.

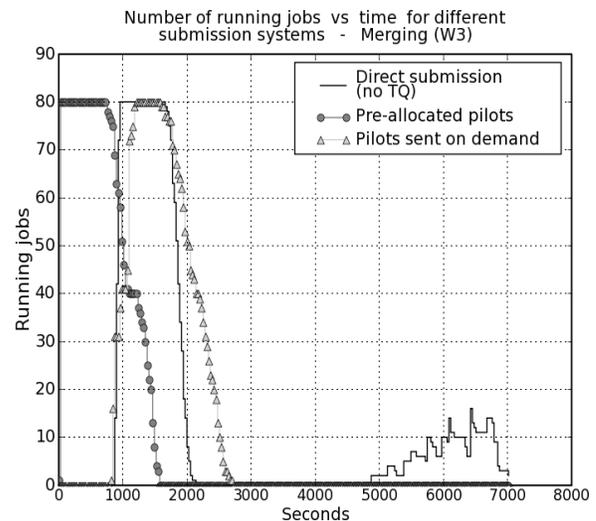

Fig. 7. Number of running jobs over time

Figure 8 shows the effect of stage-in delays on job execution times. The stage-in time in these experiments is the time a job spends in accessing and reading a file from an SE for processing. In CMS, files are read and processed directly from the SE and these files are not downloaded to a worker node. The plot shows that the system with the pilot cache provides much better data access times especially under worse SE conditions (d3f3) as a smaller number of requests are sent to the SE, with an increasing number of datasets being available in the local caches. In the normal SE condition (d1f1), the cache does not offer significant advantage over the *no-cache* approach as the SE has low latencies and can serve the data requests almost as fast as is expected from the pilot cache. There is a minimal effect of higher delays and higher failure rates i.e. d2f2, d3f1, d3f3 on the stage-in time for the cache based system because only a few files are read from the SE. On the contrary, with an increase in the delay and the failure rate, the stage-in time significantly increases for the *no-cache* approach because all the files are read from the SE. It is important to note here that by using the cache, I/O requests to SE are decreased; therefore, its use may also minimize the deterioration of the SE conditions in the first place.

From Figure 8, it is clear that the pilot cache mechanism positively affects the execution time of a real job by reducing the data access latencies. Since the jobs are inter-dependent in a workflow, this result should also reduce the turnaround time of a complete workflow as shown in Figure 9. From this figure, it is clear that the cache approach provides better workflow turnaround time than the *no-cache* approach. An interesting fact to note here is that an increase in the failure rate has a more prominent effect on the turnaround time compared to an increase in the delay factor. This is due to the fact that a failure in reading or writing data to an SE causes a job to fail which triggers the resubmission of a job, and causes an additional delay of resubmission and re-execution of a job. We know that a job, in a workflow, cannot be ready for execution until its predecessor job has been completed. Since the failure of one job delays the start of its dependent job, it increases the turnaround time of a workflow.

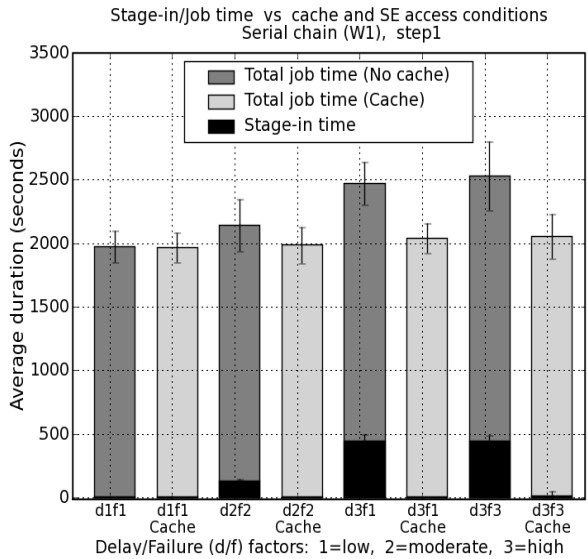

Fig. 8. Effect of data cache on stage-in time under different SE conditions

In the case of the cache based approach for W1, jobs mostly read the required files from the cache which reduces the data access latencies and the failures during the stage-in time. However, the failures at stage-out (writing data back to an SE) can lead to long workflow turnaround times. As a result, the turnaround time of W1 under d3f1 (highest delay, low failure rate) condition for both the approaches, the cache and the *no-cache*, is less than d2f2 (high delay, high failure rate) and d3f3 (highest delay, highest failure rate).

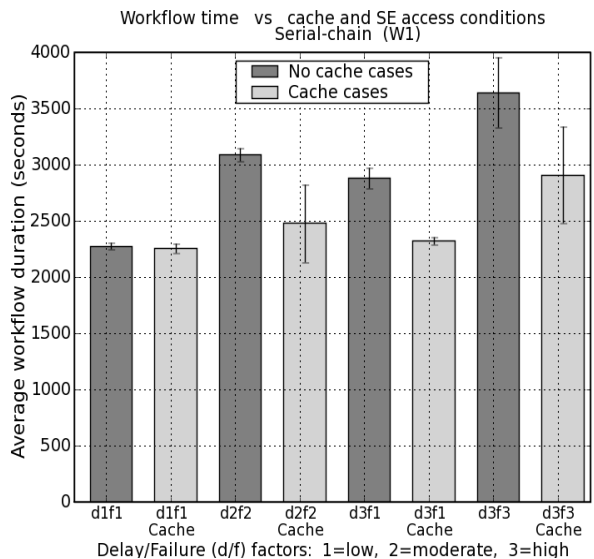

Fig. 9. Workflow turnaround time under different SE conditions

Figures 10 and 11 show the same behaviour as discussed in Figures 8 and 9 but with different workflows under worse SE conditions (d3f3). We can see that the cache mechanism performs much better for a workflow where the jobs show one-to-one dependency, i.e. the W1 workflow, because the jobs from step1 can be efficiently scheduled to the pilot jobs that hold the results of the jobs from step0.

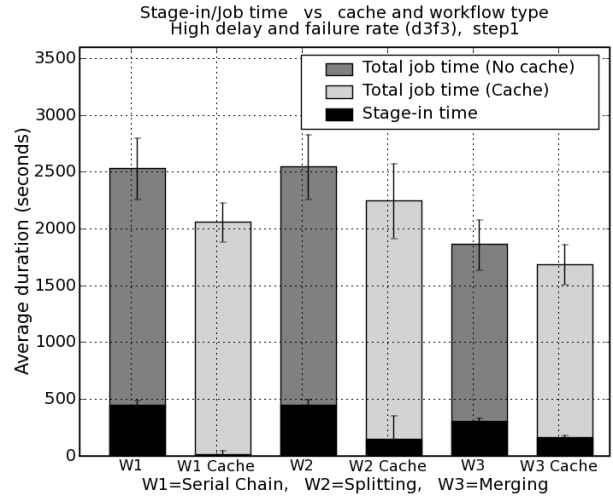

Fig. 10. Effect of cache on job execution for different workflow types

For the workflows W2 and W3, the cache hit rate is less than the one in W1 because the jobs from step1 may be forced to read from two different pilots (in case of W3) or two jobs in different pilots may read from a single pilot that is holding the data (in case of W2). On average, for all the three workflows, the system with the pilot cache behaves better than the one without it. However, the cache hits can be further increased in case of W2 and W3 if a system with a global cache is used. The global cache means that the pilot jobs can share their data across WNs in a site.

Figure 11 depicts the cache impact on the turnaround times of different workflows under worse SE conditions. It is clear that the pilots with data cache help in improving the workflow execution time when the storage resources are in the stressed conditions.

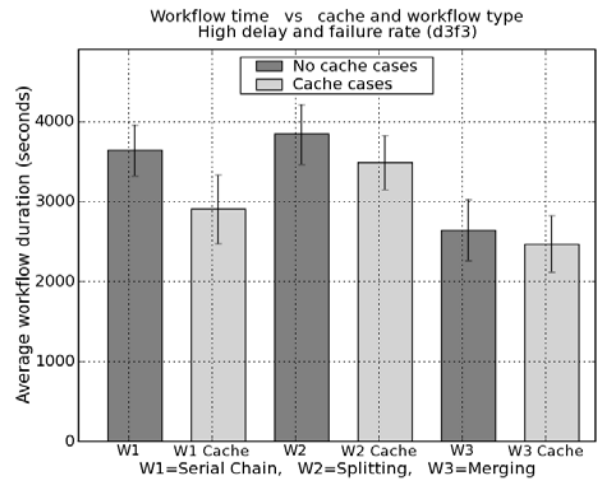

Fig. 11. Workflow turnaround times for different workflow types

In this paper, we have discussed the single pilot cache (per-pilot cache) and the cache sharing (cache-per-host) among the pilot jobs on a worker node. A caching scheme is measured on the basis of its responsiveness to the data access needs, mostly measured in terms of the cache hit ratio and the byte ratio. The cache hit ratio is the percentage of the data accesses that were found in the cache.

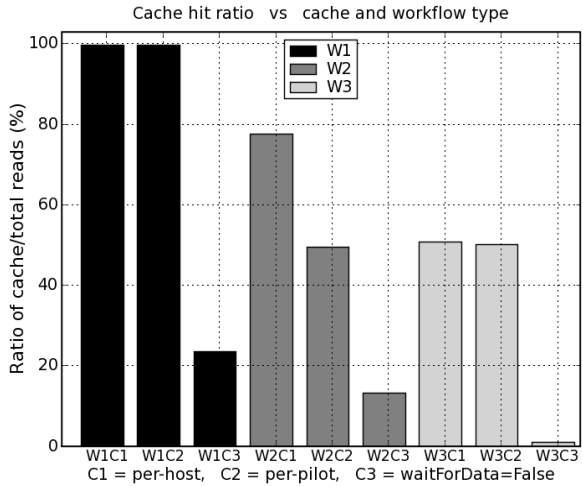

Fig. 12. Cache hit ratio

The illustration in Figure 12 depicts the cache hit ratio for different caching combinations and different types of workflows. For the serial chain workflow (W1), the single pilot cache and the share cache with the waitForData policy, as explained in Section IV-E, yield equal hit rates because the jobs have a one-to-one data dependency and they are scheduled to those pilot jobs that are holding the required files in their caches. When the waitForData policy is not in use, the cache hit rate is severely reduced because the *TaskQueue* does not wait for the pilot jobs with the required data to request the real job. Consequently, a real job is scheduled to a pilot job that may not be holding the required files in its cache, thus, it may reduce the cache hit ratio. In the absence of a global cache, the waitForData approach appears to be fundamental to achieve a good cache hit ratio because it emphasises more on data availability in the job scheduling process to increase the probability of cache hit.

In Figure 12, we can see that the efficacy of the *cache-per-host* (C1) is more prominent for the splitting workflow (W2) and exhibits a marginally better cache hit ratio than the single pilot cache (C2) for the merging workflow (W3). As we know the step0 jobs can be scheduled to any pilot job because they do not have any data dependencies. Therefore, the outputs produced by these jobs are available randomly among all the pilot jobs. For W2, each step0 job produces two output files and each step1 job requires a single file as its input. In case of the *single-pilot-cache*, we may achieve 50% cache hits at the most. However, in case of the *cache-per-host*, the reason for a better cache hit ratio is due to the possibility that two jobs might be scheduled to two pilot jobs on the same WN where the required data was produced by a step0 job. Due to the cache sharing, the jobs can discover files available in some other pilot's cache, thus increasing the cache hit ratio.

In the case of W3, each job in step1 requires at least two input files produced by two different step0 jobs, which were executed by two different pilot jobs. It may be possible that those pilot jobs are running either on two different WNs or on the same WN. In any case, with the *single-pilot-cache*, 50% cache hits might be achievable. Since the pilots cannot share caches, each merge job can find at least one file from the pilot's cache. However, in the case of the *cache-per-host*, it might be possible that multiple required files are held by multiple pilot jobs running on the same WN (if the corresponding step0 jobs were executed on this WN). In this case, the scheduling mechanism may schedule the merge job to a pilot running on this WN. However, since the files produced by the step0 jobs are available randomly among the pilot jobs, the probability of finding two required files on the same WN is very low and therefore the cache hits ratio for the *cache-per-host* is only slightly higher than the *single-pilot-cache* for W3.

## VI. CONCLUSION AND FUTURE DIRECTIONS

In this paper, we have proposed a pilot job with data cache approach to improving workflow scheduling and execution times for the CMS Tier 0 analysis workflows. This approach makes use of caching techniques to reduce data access latencies which have a major impact on the overall workflow turn-around time. We also discussed the impact of the proposed approach on the lifecycle of an individual workflow actor and also on an entire workflow. The results have shown that the proposed approach can significantly reduce the overall execution time of a workflow by reducing the scheduling and data access latencies. The reduction in these latencies is very important for latency critical systems such as the CMS Tier0.

Currently we have tested this framework using the Tier0 workflow at the CMS Tier0 infrastructure and at CIEMAT through simulated experiments. In future, we intend to test this at the CERN Tier0 site at full scale and then expand its deployment and will study its feasibility on a wider scale, ideally across the whole WLCG. In future, we aim to extend the pilot jobs based approach to address the job priorities, which are assigned by the users, in the scheduling process. We also aim to implement an intelligent approach that can cooperatively and efficiently distribute the jobs over multiple sites with minimum latencies. In the current implementation, the pilot jobs can share their caches on the same WN. However, it will be quite invaluable for the improved Grid operations to investigate the effects of the cache sharing among the pilot jobs running on different WNs within a site and even across the sites.

For a better cache replacement policy, we will investigate the effects of variants of LRU and LFU and will explore how in-memory databases can play a role in improving the cache access times when the number of read requests are scaled up to thousands as is the case in the Grid infrastructures such as

WLCG. In this paper, we have assumed that a pilot job can run for an unlimited time but this might not be the case in the production Grid infrastructures. Therefore, in future, investigations will also be made to study the impact of the pilot lifetime on workflow execution.